# Student created video content for assessment, experiential learning, and service

Ben Graul[1], Matthew L. Rollins[1], Nathan Powers[1], Dennis Della Corte[1,*]
[1]Department of Physics and Astronomy, Brigham Young University, Provo, Utah
[*]Dennis.DellaCorte@byu.edu

This paper outlines the feasibility of replacing a midterm assessment with a student-created digital media project. We examine the benefits of a peer review process for student-created pedagogy, the effectiveness of the student-generated digital media, and self-evaluations as a replacement for accepted forms of assessment in terms of: learning, assessment, and service provided. The outlined project also facilitates the creation of a video bank easily accessible for professors and students alike in introductory physics courses using the OpenStax College Physics textbook.[1] We conclude that this non-traditional teaching and assessment model is both effective and timely as more college courses are being taught digitally and/or asynchronously.

## Introduction

In recent years, there has been an increasing interest in developing teaching methodologies beyond the traditional lecture-based classroom. Technology has drastically challenged the assumed 'optimal' nature of the classroom, and various instructors have rigorously examined traditional models of teaching and learning for their efficacy.[2-7] The importance of using effective technology has been largely an intellectual exercise for years; while it is logical to desire more effective student outcomes and learning, the necessity of adapting classrooms was not urgent. However, as the global COVID-19 pandemic disrupted traditional classroom instruction across the world, delivering effective pedagogy in a non-traditional manner rapidly evolved from a possible course of action to a pressing necessity.

Currently, instructors are struggling to digitize courses as the certainty of in-person learning is in question for the foreseeable future. However, delivering content is just the first step in fostering an effective online learning environment. Students must also be prepared to effectively use skills and knowledge learned after they have completed their education. This preparation includes effective peer interactions, demonstrating understanding, digital fluency, and consideration for their ability to help others through their efforts in learning material. In light of this, we developed a protocol that encourages students to create a video teaching introductory physics concepts - leveraging digital technologies and peer feedback. This process adds the benefit of democratizing the creation of effective didactic content to provide future instructors with ready-made videos, taught from a learner's perspective.

Using this protocol provides a semester-long project to replace a single midterm assessment score. Since non-traditional classroom models are necessary as distance learning becomes more common, this assessment provides significant flexibility in internet-centric courses. This paper will begin by considering the conceptual foundation for a learner-developed digital media approach, then presents one suggestion for implementing such a strategy. It will conclude with a summary of best practices and recommendations if the methodology is to be adopted. We investigate the benefits of this strategy with respect to student learning, efficacy of assessment, and service provided to peers and the community.

*Learning*

Introductory physics courses at the university level typically have an extremely high student-to-instructor ratio. For courses with hundreds of students (344 in our case), it is not feasible for an instructor to be able to deliver detailed, personal feedback for each project submitted in a timely manner. Allowing the students to review their peers' work and deliver feedback can facilitate the use of creative projects (such as videos) to reinforce learning. However, this learning is not limited to physics-related topics.

The process developed relies on the efficacy of using what has been termed "learner-generated digital media" to deepen conceptual understanding of introductory physics topics.[8, 9] Using an asynchronous method of assessment which encourages both mastery of the subject and fluency in digital media benefits the student beyond standard memorization or concept application.[8, 10, 11] Technological fluency is an undisputed necessity in the contemporary market,[11] and trends in technology and policy imply that effective presentation skills within the digital space will appreciate in value. Creating videos as a type of digital authorship allows students to synthesize material and develop skills that will be valuable as they progress throughout their careers.[12] As such, providing students with opportunities to use technology in an approximation of the type of presentation skills that will likely be required in webinars, didactic material, job interviews, and other digital communications of complex information becomes as necessary as teaching effective writing strategies.[9, 13, 14]

Students developing and delivering feedback has benefits beyond ensuring mastery of the material. This peer review results in timely feedback for the student, but more importantly leads to improvements in written communication and interpersonal skills.[15] Upon receiving feedback, the student is prompted to reflect upon their understanding of the material as they decide what aspects to implement in later versions.[13] This often prompts the student to review the relevant material and seek answers to new questions that arise due to feedback.

*Assessment*

Large projects typically require a significant amount of time to adequately grade and provide feedback. However, using peer review for grading videos frees up time for the instructor and teaching assistants. The students' lack of expertise in subject matter and grading ability can be overcome by effective preparation. Students who receive detailed rubrics and brief training can assign grades that do not significantly differ from expert graders and deliver substantive feedback.[16] Additionally, students who are in the process of learning concepts also have insight into which aspects of a topic are difficult to grasp, whereas experts might easily overlook such difficulties or underestimate their complexity for a novice.[17]

*Service*

There is additional benefit in this method of evaluation due to student appreciation of their work's importance. Creating a video that could be used to instruct future cohorts of students can be a motivating factor for emphasizing the quality of the video created.[18] For example, multiple students expressed that they hoped their video would be helpful for other introductory physics students in a final self-reflection statement. The act of creating a project that will be viewed and used beyond a single review for grading provides extrinsic motivation for students to create higher-quality projects.

## Methods

344 students were tasked with creating an educational video on one of 70 General Physics 2 related topics over the course of the semester. While a few media types were mentioned in class as examples, no requirements or expectations were made explicit in regard to the media style. At the midpoint of the semester, students submitted a first video to receive feedback from their peers.

Each student was assigned five videos of a topic to review and given a form to complete as they delivered feedback. Students then implemented improvements to their video and submitted a second version for an additional peer review. Finally, students submitted an assessment of their own work based on reviews and grades received, which included a self-assigned grade.

In order to compare the value and legitimacy of the peer reviews, a rubric was constructed from which teaching assistants could provide a graded review for a randomly selected portion of the videos. This rubric delineated point values given for specific aspects of the video's presentation and content. Table 1 compares our methodology with similar studies examining the best practices for implementing student-made media projects within introductory physics courses.

**Analysis**

Our analysis demonstrates key points which both support and occasionally challenge best practices established in the literature. As shown in Figure 1-a, peers skew the grades when they recognize the grade they are giving as final. This suggests that training and a clear rubric are necessary for peer reviews to be properly used as an assessment tool. Figure 1-b shows an overall improvement in video quality from the second submission, suggesting that an iterative process and peer review contributes to learning. Similarly, the grade distribution closely resembles that of the first midterm from the course. As such, a similar video project could be appropriate in place of a midterm assessment. Figure 1-c expresses both a majority positive experience with the peer review process as well as perceived experiential learning. Finally, the wide spread of media types and teaching strategies observed and shown in Figure 1-d imply that limited instruction can nurture a creative environment.

| Project Component | Comparable Methods | Our Method | Observation | Recommended Action |
|---|---|---|---|---|
| **Motivation for Project** | Focus on skills used beyond classroom[8, 9, 11, 19, 20] | Did not explicitly mention value of process outside of classroom | Majority of students mention benefits beyond the classroom | Emphasize beneficial writing, presentation, research, and multimedia skills to be developed |
| **Grading Criteria** | Create score-based Rubric[15] | Rubric based on qualitative assessment | TA grades utilizing quantitative rubric are more evenly distributed for final submissions | Quantitative rubric developed for project should be utilized by students |
| **Project Design** | Project should be an iterative process (e.g. storyboard before production) [11, 16, 18, 21] | Students turned in rough draft and final draft | Second draft was typically higher quality | Have at least two versions with feedback delivered on each |
| **Teaching Strategies and Media Usage** | Identify proper digital media types to utilize[9, 11] | Gave limited example technologies | Not specifying media types facilitated diverse teaching strategies | Example types can be provided, but creativity should be encouraged |
| **Training of content producers** | Students trained extensively about Digital Media creation.[10] | Students teach themselves about creating digital media | Training effect observed due to repetition within iterative protocol | Using iterative process can replace the need for training |
| **Grading Training** | Focused student training on grading[16] | Students grade based on qualitative rubric | Lack of training leads to highly skewed peer grades | Give example grading process using quantitative rubric |
| **Follow-up Activity** | Opportunity for self-reflection[22] | Provided chance to write self-reflection and assign a personal grade | Students made meaningful changes to projects after reflecting on quality of work following rough draft | Provide opportunity to reflect on work at midpoint and end of semester |

**TABLE 1: Comparison of essential video experiment components between literature and presented project. Summary of findings and suggested recommendations.**

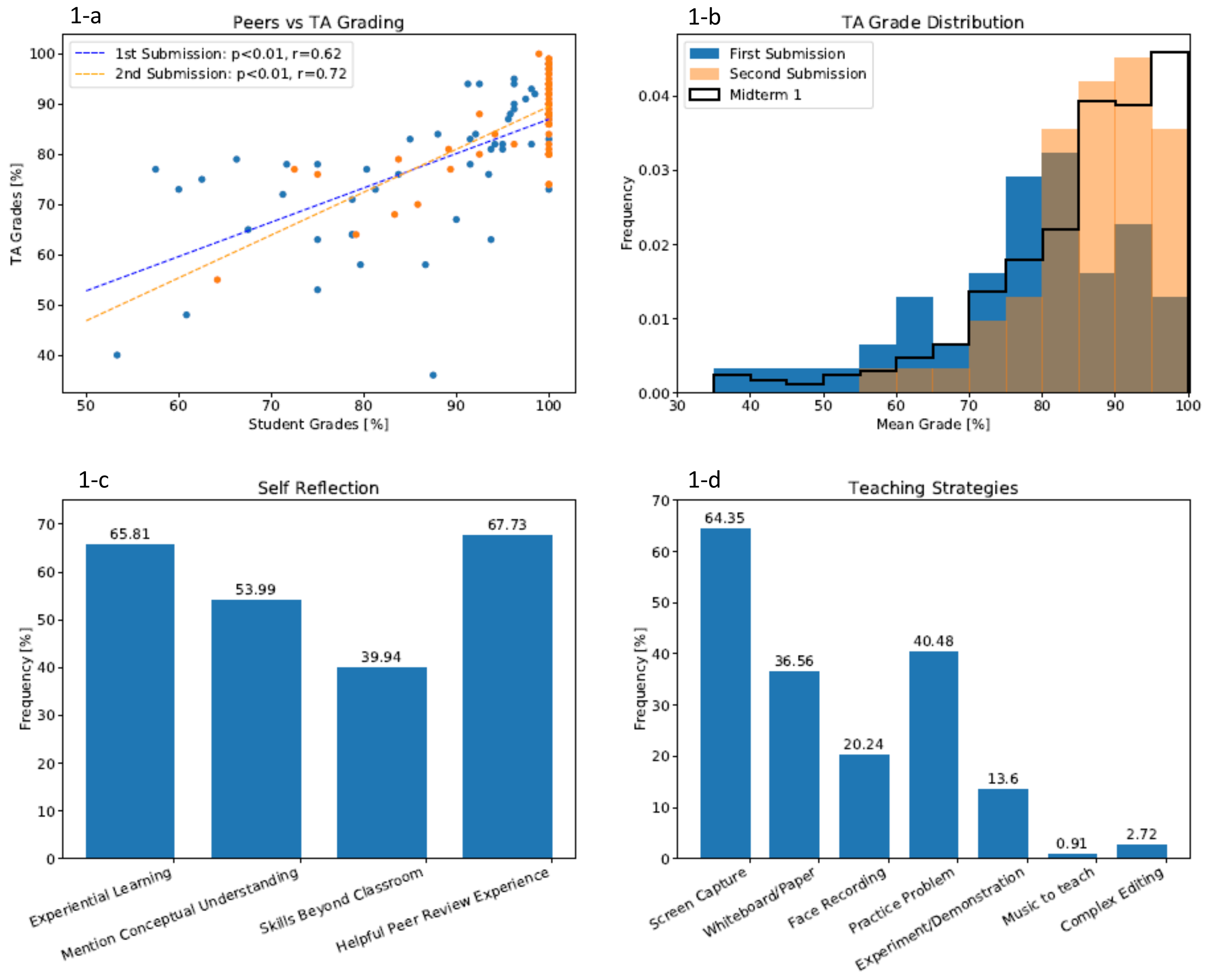

**Figure 1: Quantitative analysis of key hypotheses. 1-a: Correlation analysis between student and Teaching Assistants (TA) grades for first and second submissions. 1-b: TA grade distribution for first and second submission compared to standard midterm 1 score distribution. 1-c: Histogram of learning outcomes as mentioned within student self-reflections. 1-d: Histogram of teaching strategies utilized within videos.**

## Discussion

*Learning*

Creating an educational video leads to unique learning opportunities. A majority of students stated that they felt they learned lessons through this exercise that they did not attain via traditional lectures or exams. Many also mentioned they felt they had developed skills that would be useful in other classes, their future jobs, and in their personal lives.

We observed a trend when students actively reflected on the review from the first draft of their video. Many students mentioned some feedback they received that led to improving their video. However, the more substantial changes students mentioned often arose from watching their peers' videos and considering how their own work compared. Within their self-reflection paper, many students suggested that reflecting on their own understanding was at least as helpful as explicitly receiving feedback.

By providing limited instruction on the type of teaching style expected, students exercised enormous creativity in conception and execution for the videos. While a majority of students created a fairly straightforward lecture-style video, others created extensive demonstrations (e.g. functioning engines or deconstructing electric guitars), wrote and starred in rap music videos, or created documentary-style presentations. Videos that included these unique aspects were appreciated by peers, and many stated that their unique approach helped them understand concepts when traditional lectures had been unable to do so.

We conclude, contrary to a previous study[23] but in agreement with another[24] that many benefits come from allowing students flexibility in identifying how they learn best, as well as participation in active learning through the creation of relevant video content.[25] Some students report higher learning outcomes from written materials, while others report a preference for didactic videos. Each provides unique benefits compared to the other, allowing students the ability to choose their preferred method of learning may lead to higher satisfaction and learning as will be investigated in a follow up experiment.

*Assessment*

Students must be motivated with an incentive to make high quality videos. This incentive will typically be a significant portion of a final grade. However, students seem to be averse to giving poor grades on a final assessment. The resulting grade inflation means that the project should not replace a final assessment, but a midterm may be acceptable. We observed a significant bias towards high scores in the grading process, especially in the self-assessment reflection papers assigned. Many students stated that although their peers had graded them poorly, the amount of work they performed was deserving of a higher grade. Additionally, the peer feedback papers often had sharp criticisms of the video but gave a very generous grade. A notable example stated that a reviewed video was likely plagiarized, but still gave a grade of C+. However, using peers instead of TAs decreases class overhead substantially. With an effective rubric and training in delivering feedback, student grades can be used effectively.

*Service*

We anticipate that future cohorts of students will leverage the best videos, indicated by positives votes on YouTube, to complement reading assignments in the class textbook.[1] It is assumed that other institutes leveraging this text will also benefit from this project. Over time, a list of effective teaching videos can be cultivated and shared between institutes and programs. Furthermore, students stated that they felt motivated to create a more detailed and interesting product because they expected other students to learn from their work. Multiple students mentioned an increased desire to learn and teach effectively due to the expectation that others would watch the video, while they often felt that an essay would be quickly forgotten or skimmed through.

## Conclusion

Assigning students to create didactic videos is a viable method of encouraging learning, service, and effectively assessing progress in introductory physics courses. Between video creation and peer review, students learn relevant physics concepts and necessary skills to succeed in future endeavors. In order to effectively implement the process, it is necessary to provide a detailed rubric, chances for students to receive and implement peer feedback, and a final reflection on the overall learning process.

## References

1. O. College, *College Physics*. (Openstax, 2012).
2. M. Kearney, Learning, Media and Technology **36** (2), 169-188 (2011).
3. I. Kosterelioglu, Universal Journal of Educational Research **4** (2), 359-369 (2016).
4. T. L. Larkin, International Journal of Engineering Pedagogy (iJEP) **4** (2), 36 (2014).
5. M. R. Laugerman and K. P. Saunders, Decision Sciences Journal of Innovative Education **17** (4), 387-404 (2019).
6. C. J. Luxford and S. L. Bretz, Chemistry Education Research and Practice **14** (2), 214-222 (2013).
7. N. Mirriahi, D. Liaqat, S. Dawson and D. Gaševic, Educational Technology Research and Development **64** (6), 1083-1106 (2016).
8. J. Reyna, J. Hanham and P. Meier, E-Learning and Digital Media **14** (6), 309-322 (2017).
9. J. Reyna, J. Hanham, P. Vlachopoulos and P. Meier, Research in Science Education (2019).
10. R. C. Choe, Z. Scuric, E. Eshkol, S. Cruser, A. Arndt, R. Cox, S. P. Toma, C. Shapiro, M. Levis-Fitzgerald, M. Levis-Fitzgerald, G. Barnes and R. H. Crosbie, CBE - Life Sciences Education **18** (4) (2019).
11. W. Nielsen, H. Georgiou, P. Jones and A. Turney, Research in Science Education (2018).
12. C. Snelson, Learning, Media and Technology **43** (3), 294-306 (2018).
13. A. S. Halim, S. A. Finkenstaedt-Quinn, L. J. Olsen, A. R. Gere and G. V. Shultz, CBE—Life Sciences Education **17** (2), ar28 (2018).
14. E. Price, F. Goldberg, S. Robinson and M. McKean, Physical Review Physics Education Research **12** (2) (2016).
15. K. J. Topping, Theory Into Practice **48** (1), 20-27 (2009).
16. S.-Y. Lin, S. S. Douglas, J. M. Aiken, C.-L. Liu, E. F. Greco, B. D. Thoms, M. D. Caballero and M. F. Schatz, arXiv preprint arXiv:1407.4714 (2014).
17. L. Rubin and C. Hebert, College Teaching **46** (1), 26-30 (2010).
18. J. Pirhonen and P. Rasi, Journal of Biological Education **51** (3), 215-227 (2017).
19. M. Gallardo-Williams, L. A. Morsch, C. Paye and M. K. Seery, Chemistry Education Research and Practice **21** (2), 488-495 (2020).
20. G. Hoban, W. Nielsen and A. Shepherd, *Student-Generated Digital Media in Science Education: Learning, Explaining and Communicating Content*. (2015).
21. J. Wiggen and D. McDonnell, Journal of College Science Teaching **46** (6), 44-49 (2017).
22. M. E. Walvoord, M. H. Hoefnagels, D. D. Gaffin, M. M. Chumchal and D. A. Long, Journal of College Science Teaching **37** (4), 66-73 (2008).
23. J. P. Abulencia, D. Silverstein and M. Vigeant, in *121st ASEE Annual Conference & Exposition* (American Society for Engineering Education, Indianapolis, IN, 2014), Vol. 24, pp. 1.
24. P. J. Stephens, World Journal of Education **7** (3), 14-20 (2017).
25. A. Weinberg and M. Thomas, International Journal of Mathematical Education in Science and Technology **49** (6), 922-943 (2018).

Supplementary Information

# Student created video content for assessment, experiential learning, and service

Graul, B. et al.

| OPENSTAX CHAPTER | Name of video | Url link |
|---|---|---|
| 18.1 | The Triboelectric Effect | https://www.youtube.com/watch?v=jODOMgydj2I&list=PL6oBCjihmcBmAsb_4rkKz5FKEUSJD4M4I&index=268&t=0s |
| 18.1 | The Triboelectric Effect | https://www.youtube.com/watch?v=-WrIp4VTMv4&list=PL6oBCjihmcBmAsb_4rkKz5FKEUSJD4M4I&index=214&t=0s |
| 18.1 | Triboelectric Effect | https://www.youtube.com/watch?v=B3UKf7HlwKk&list=PL6oBCjihmcBmAsb_4rkKz5FKEUSJD4M4I&index=130&t=0s |
| 18.1 | The Triboelectric Effect | https://www.youtube.com/watch?v=7aGi_RsG4dI&list=PL6oBCjihmcBmAsb_4rkKz5FKEUSJD4M4I&index=188&t=0s |
| 18.1 | Triboelectric Effect | https://www.youtube.com/watch?v=iwyrdO8NPUo&list=PL6oBCjihmcBmAsb_4rkKz5FKEUSJD4M4I&index=344&t=0s |
| 18.3 | Coulomb's Law | https://www.youtube.com/watch?v=B4XGgiAYioM&list=PL6oBCjihmcBmAsb_4rkKz5FKEUSJD4M4I&index=282&t=0s |
| 18.3 | Coulomb's Law | https://www.youtube.com/watch?v=M8A7PJZqIWo&list=PL6oBCjihmcBmAsb_4rkKz5FKEUSJD4M4I&index=334&t=0s |
| 18.3 | Coulomb's Law | https://www.youtube.com/watch?v=aDDo_1f4Bt8&list=PL6oBCjihmcBmAsb_4rkKz5FKEUSJ%20D4M4I&index=307&t=0s |
| 18.3 | Coulomb's Law | https://www.youtube.com/watch?v=bS7VS-DKqiY&list=PL6oBCjihmcBmAsb_4rkKz5FKEUSJD4M4I&index=295&t=0s |
| 18.4-18.5 | Electric Field and Field Lines | https://www.youtube.com/watch?v=y-0Ax27qCt4&list=PL6oBCjihmcBmAsb_4rkKz5FKEUSJD4M4I&index=258&t=0s |
| 18.4-18.5 | Electric Field Lines | https://www.youtube.com/watch?v=2NhBzA8MtKI&list=PL6oBCjihmcBmAsb_4rkKz5FKEUSJD4M4I&index=298&t=0s |
| 18.4-18.5 | Electric Fields and Electric Field Lines | https://www.youtube.com/watch?v=nPpgRfz_K7E&list=PL6oBCjihmcBmAsb_4rkKz5FKEUSJD4M4I&index=160&t=0s |
| 18.4-18.5 | Electric Field Lines | https://www.youtube.com/watch?v=PdQyF_4u8tM&list=PL6oBCjihmcBmAsb_4r |
| 18.4-18.5 | Electric Fields and Electric Field Lines | https://www.youtube.com/watch?v=amFrhQ9GnLo&list=PL6oBCjihmcBmAsb_4rkKz5FK |
| 18.6 | Connection between Electric Force and Electric Field | https://www.youtube.com/watch?v=yFAzVIv_d3k&list=PL6oBCjihmcBmAsb_4rkKz5FKEUSJD4M4I&index=262&t=0s |
| 18.6 | Relationship Between Electric Force and Field | https://www.youtube.com/watch?v=WOEF7t2TN-A&list=PL6oBCjihmcBmAsb_4rkKz5FKEUSJD4M4I&index=172&t=0s |
| 18.6 | Electric Field & Force | https://www.youtube.com/watch?v=u47LZKrqIAs&list=PL6oBCjihmcBmAsb_4rkKz5FKE |
| 18.6 | Electric Force Vs Electric Field. | https://www.youtube.com/watch?v=hir2huQ0gJA&list=PL6oBCjihmcBmAsb_4rkKz5FKEUSJD4M4I&index=216&t=0s |
| 18.7 | Conductors, insulators, and charging | https://www.youtube.com/watch?v=EDwCJAJtivc&list=PL6oBCjihmcBmAsb_4rkKz5FKEUSJD4M4I&index=330&t=0s |
| 18.7 | Conductors Insulators and Charging | https://www.youtube.com/watch?v=gC_9U_2d6V4&list=PL6oBCjihmcBmAsb_4rkKz5FKEUSJD4M4I&index=292&t=0s |
| 18.7 | Charges Conductors and Insulators | https://www.youtube.com/watch?v=2o1alzxVa3I&list=PL6oBCjihmcBmAsb_4rkKz5FKEUSJD4M |
| 18.7 | Conductors and Insulators | https://www.youtube.com/watch?v=MuCP4dNhjl4&list=PL6oBCjihmcBmAsb_4rkKz5FKE |
| 18.7 | Conductors Electric Field | https://www.youtube.com/watch?v=SCpiE_9Oiw8&list=PL6oBCjihmcBmAsb_4rkKz5FKEUSJD4M4I&index=199&t=0s |
| 18.7 | Conductors & ElectricFields | https://www.youtube.com/watch?v=ldfaVCsY8CI&list=PL6oBCjihmcBmAsb_4rkKz5FKEUSJD4M4I&index=4&t=0s |
| 18.7 | Conductors And Electric Fields | https://www.youtube.com/watch?v=tYPf0Pjf0QI&list=PL6oBCjihmcBmAsb_4rkKz5FKEUSJD4M4I&index=228&t=0s |
| 18.7 | Conductors & Electric Field | https://www.youtube.com/watch?v=WNMnj8dm6Hg&list=PL6oBCjihmcBmAsb_4rkKz |
| 18.7 | ELECTRICFIELDS AND CONDUCTORS | https://www.youtube.com/watch?v=vstQJ8WTokQ&list=PL6oBCjihmcBmAsb_4rkKz |
| 18.8 | Van De Graff | https://www.youtube.com/watch?v=EAgMW7zbn3M&list=PL6oBCjihmcBmAsb_4rkKz5FKEUSJD4M4I&index=324 |
| 18.8 | Van De Graff | https://www.youtube.com/watch?v=1D8MZA9Pv4o&list=PL6oBCjihmcBmAsb_4rkKz5FKEUSJD4M4I&index=183&t=0s |
| 18.8 | Van De Graaff | https://www.youtube.com/watch?v=g3JOiOR_knc&list=PL6oBCjihmcBmAsb_4rkKz5FKE |
| 18.8 | VAN DE GRAAFF | https://www.youtube.com/watch?v=OL_PRM3bGik&list=PL6oBCjihmcBmAsb_4rkKz5FKEUSJD4M4I&index=164&t=0s |
| 18.8 | Van de Graaff Generators | https://www.youtube.com/watch?v=ZuoVhs5MJhE&list=PL6oBCjihmcBmAsb_4rkKz5FKEUSJD4M4I&index=90&t=0s |
| 19.1 | Electric Potential | https://www.youtube.com/watch?v=i1Ew3CYiXh4&list=PL6oBCjihmcBmAsb_4rkKz5FKEUSJD4M4I&index=47 |
| 19.1 | Electric Potential | https://www.youtube.com/watch?v=D4Oj6lxFrvM&list=PL6oBCjihmcBmAsb_4rkKz5FKEUSJD4M4I&index=318&t=0s |
| 19.1 | Electric Potential | https://www.youtube.com/watch?v=952HU33LoB0&list=PL6oBCjihmcBmAsb_4rkKz5FKEUSJD4M4I&index=308&t=0s |
| 19.2 | Relationship Electric Potential & Field | https://www.youtube.com/watch?v=YXPDF_xGS50&list=PL6oBCjihmcBmAsb_4rkKz5FKEUSJD4M4I&index=50&t=0s |
| 19.2 | RELATIONSHIP BETWEEN ELECTRIC POTENTIAL AND ELECTRIC FIELD | https://www.youtube.com/watch?v=vBfr9SZsNMI&list=PL6oBCjihmcBmAsb_4rkKz5FKEUSJD4M4I&index=119&t=0s |
| 19.2 | Relationship Electric Field Electric Potential | https://www.youtube.com/watch?v=CTHTzYhQxl0&list=PL6oBCjihmcBmAsb_4rkKz5FKEUSJD4M4I&index=257&t=0s |
| 19.2 | Electric field & ElectricPotential | https://www.youtube.com/watch?v=EyLS8tg8plQ&list=PL6oBCjihmcBmAsb_4rkKz5FKEUSJD4M4I&index=284&t=0s |
| 19.2 | Electric Fields and Electric Potential | https://www.youtube.com/watch?v=rwgzv_Ejesg&list=PL6oBCjihmcBmAsb_4rkKz5FKEUSJD4M4I&index=238&t=0s |
| 19.4 | Equipotentials and Field Lines | https://www.youtube.com/watch?v=PsqU_Ke1YUA&list=PL6oBCjihmcBmAsb_4rkKz5FKEUSJD4M4I&index=338&t=0s |
| 19.4 | Equipotential and Field Lines | https://www.youtube.com/watch?v=RpH3hrr4Vms&list=PL6oBCjihmcBmAsb_4rkKz5FKEUSJD4M4I&index=210&t=0s |
| 19.4 | Equipotential Lines | https://www.youtube.com/watch?v=nGoWI8HaYSQ&list=PL6oBCjihmcBmAsb_4rkKz5FKEUSJD4M4I&index=79&t=0s |
| 19.4 | Equipotentials and Field Lines | https://www.youtube.com/watch?v=aBBItYaeLoc&list=PL6oBCjihmcBmAsb_4rkKz |
| 19.6 | Capacitors In Series And Parallel | https://www.youtube.com/watch?v=h-1FH3o-BVY&list=PL6oBCjihmcBmAsb_4rkKz5FKEUSJD4M4I&index=340&t=0s |
| 19.6 | Capacitors In Series And In Parallel | https://www.youtube.com/watch?v=ylOYM_0NXp4&list=PL6oBCjihmcBmAsb_4rkKz5FKEUSJD4M4I&index=62 |
| 19.6 | Capacitors In Series And Parallel | https://www.youtube.com/watch?v=92xZZuqfQmk&list=PL6oBCjihmcBmAsb_4rkKz5FKEUSJD4M4I&index=108&t=0s |
| 19.6 | Capacitors in Series and in Parallel | https://www.youtube.com/watch?v=0PidplJavUY&list=PL6oBCjihmcBmAsb_4rkKz5FKEUSJD4M4I&index=277&t=0s |
| 19.6 | Capacitors In Series And In Parallel | https://www.youtube.com/watch?v=Q8SUaAaXGkY&list=PL6oBCjihmcBmAsb_4rkKz5FKEUSJD4M4I&index=264&t=0s |
| 20.1 | Electric Current | https://www.youtube.com/watch?v=_x7MaBzRGkM&list=PL6oBCjihmcBmAsb_4rkKz |
| 20.1 | Electric Currents | https://www.youtube.com/watch?v=4tk83VU1XU0&list=PL6oBCjihmcBmAsb_4rkKz5FK |
| 20.1 | Electric Current | https://www.youtube.com/watch?v=OBhe7FljTeg&list=PL6oBCjihmcBmAsb_4rkKz5FKEUSJD4M4I&index=117&t=0s |
| 20.1 | Electric Current | https://www.youtube.com/watch?v=OFMRcZxxT_A&list=PL6oBCjihmcBmAsb_4rkKz |
| 20.1 | Speed of current | https://www.youtube.com/watch?v=w8HH6rP39M0&list=PL6oBCjihmcBmAsb_4rkKz5FKEUSJD4M4I&index=230&t=0s |
| 20.1 | Speed of Current | https://www.youtube.com/watch?v=uTEbtE7U9ME&feature=youtu.be |
| 20.1 | How Fast Does Current Travel | https://www.youtube.com/watch?v=KEElzJeAE7s&list=PL6oBCjihmcBmAsb_4rkKz5FKEUSJD4M4I&index=290&t=1s |
| 20.1 | How Fast Does Current Travel? | https://www.youtube.com/watch?v=hB8NDdfxcb8&list=PL6oBCjihmcBmAsb_4rkKz5FKEUSJD4M4I&index=316&t=0s |
| 20.2 | Ohms Law | https://www.youtube.com/watch?v=9o-5ospiDuk&list=PL6oBCjihmcBmAsb_4rkKz5FKEUSJD4M4I&index=136&t=0s |
| 20.2 | Ohms Law | https://www.youtube.com/watch?v=hzlxsfEM5Ec&list=PL6oBCjihmcBmAsb_4rkKz5FKEUSJD4M4I&index=168 |
| 20.2 | Ohm’s Law | https://www.youtube.com/watch?v=VnqMrCrz9xU&list=PL6oBCjihmcBmAsb_4rkKz5FKEUSJD4M4I&index=233&t=0s |
| 20.2 | OHMSLAW. | https://www.youtube.com/watch?v=snYPXst9jWM&list=PL6oBCjihmcBmAsb_4rkKz5FKEUSJD4 |
| 20.3 | Resistance And Superconductivity | https://www.youtube.com/watch?v=vjX3Du1aCIk&list=PL6oBCjihmcBmAsb_4rkKz5FKEUSJD4M4I&index=241&t=0s |
| 20.3 | Temp Dependence of Resistors and Superconductivity | https://www.youtube.com/watch?v=otp8VLX0tEU&list=PL6oBCjihmcBmAsb_4rkKz5FKEUSJD4M4I&index=197&t=0s |
| 20.4 | Power in Circuits | https://www.youtube.com/watch?v=XBUdtWjGM10&list=PL6oBCjihmcBmAsb_4rkKz5FKEUSJD4M4I&index=103 |
| 20.4 | Power in Electric Circuits | https://www.youtube.com/watch?v=tYTG-0dYNO4&list=PL6oBCjihmcBmAsb_4rkKz5FKEUSJD4M4I&index=177&t=0s |
| 20.4 | Power In Electric Circuits | https://www.youtube.com/watch?v=QoKyHmp5RjE&list=PL6oBCjihmcBmAsb_4rkKz5FKEUSJD4M4I&index=38&t=0s |
| 20.4 | Power in circuits | https://www.youtube.com/watch?v=axvVdvL2YWE&list=PL6oBCjihmcBmAsb_4rkKz5FKEUSJD4M4I&index=212&t=0s |
| 20.5 | AC vs. DC | https://www.youtube.com/watch?v=mtTWKavTuwA&list=PL6oBCjihmcBmAsb_4 |
| 20.5 | The difference between AC & DC | https://www.youtube.com/watch?v=9GO8j8qfoCE&list=PL6oBCjihmcBmAsb_4rkKz5FKEUSJD4M4I&index=190&t=0s |
| 20.5 | DIFFERENCE BETWEEN AC & DC | https://www.youtube.com/watch?v=sAbqe2IixYI |
| 20.5 | Alternating and Direct Current | https://www.youtube.com/watch?v=-Hk4qbDvIg0&list=PL6oBCjihmcBmAsb_4rkKz5FKEUSJD4M4I&index=201&t=0s |
| 20.5 | RMS Values | https://www.youtube.com/watch?v=fKcdTVLUsWs&list=PL6oBCjihmcBmAsb_4rkKz5FKEUSJD4M4I&index=226&t=0s |
| 20.5 | RMS values | https://www.youtube.com/watch?v=j9__2NLTEgM&list=PL6oBCjihmcBmAsb_4rkKz5FKEUSJD4M4I&index=168&t=0s |
| 20.5 | RMS values | https://www.youtube.com/watch?v=iX6pSdPir1w&list=PL6oBCjihmcBmAsb_4rkKz5FKEUSJD4M4I&index=176&t=0s |
| 20.5 | RMS Values. | https://www.youtube.com/watch?v=65WtKN8QD3Q&list=PL6oBCjihmcBmAsb_4rkKz5FKEUSJD4M4I&index=58&t=4s |
| 21.1 | Resistors in Series and Parallel | https://www.youtube.com/watch?v=PUn4YB77fWI&list=PL6oBCjihmcBmAsb_4rkKz5FKEUSJD4M4I&index=68&t=0s |
| 21.1 | Resistors in Parallel and Series | https://www.youtube.com/watch?v=SAgCEs-c4X8&list=PL6oBCjihmcBmAsb_4rkKz5FKEUSJD4M4I&index=42&t=0s |
| 21.1 | Resistors | https://www.youtube.com/watch?v=P83N_tLyD2s&list=PL6oBCjihmcBmAsb_4rkKz5FKEUSJD4M4I&index=47&t=0s |
| 21.1 | Resistors In Series And Parallel | https://www.youtube.com/watch?v=g4Hfz8AhZCU&list=PL6oBCjihmcBmAsb_4rkKz5FKEUSJD4M4I&index=123&t=0s |
| 21.2 | EMF & Terminal Voltage | https://www.youtube.com/watch?v=iDZk-M4PP1g&list=PL6oBCjihmcBmAsb_4rkKz5FKEUSJD4M4I&index=320&t=0s |
| 21.2 | EMF and Terminal Voltage | https://www.youtube.com/watch?v=GxKelkIm3No&list=PL6oBCjihmcBmAsb_4rkKz5FKEUSJD4M4I&index=109&t=0s |
| 21.2 | EMF and Terminal Voltage | https://www.youtube.com/watch?v=xMY6zXll8Gk&list=PL6oBCjihmcBmAsb_4rkKz5FKEUSJD4M4I&index=135&t=0s |
| 21.2 | EMF and Terminal Voltage | https://www.youtube.com/watch?v=0JmRcPNAJyY&list=PL6oBCjihmcBmAsb4rkKz5FKEUSJD4M4I&index=191&t=0s |
| 21.2 | EMF and Terminal Voltage | https://www.youtube.com/watch?v=Heq0GxfTSZo&list=PL6oBCjihmcBmAsb_4rkKz5FKEUSJD4M4I&index=62&t=0s |
| 21.3 | Kirchhoff's Laws | https://www.youtube.com/watch?v=NIyie0eM14Q&list=PL6oBCjihmcBmAsb_4rkKz5FKEUSJD4M4I&index=209&t=0s |
| 21.3 | Kirschoffs Rules | https://www.youtube.com/watch?v=Fe9liw998V4&list=PL6oBCjihmcBmAsb_4rkKz5FKEUSJD4M4I&index=304&t=0s |
| 21.3 | Kirchhoffs Laws | https://www.youtube.com/watch?v=M2_VQmO5ztg&list=PL6oBCjihmcBmAsb_4rkKz5F |
| 21.6 | RC circuits | https://www.youtube.com/watch?v=ED7oiz3ZmdY&list=PL6oBCjihmcBmAsb_4rkKz5FKEUSJD4M4I&index=64 |
| 21.6 | RC Circuits - Rap | https://www.youtube.com/watch?v=Nzey8HP_ciQ&list=PL6oBCjihmcBmAsb_4rkKz5FKEUSJD4M4I&index=296&t=0s |
| 21.6 | RC Circuits | https://www.youtube.com/watch?v=Y-3QQaqU6eM&list=PL6oBCjihmcBmAsb_4rkKz5FKEUSJD4M4I&index=319&t=0s |
| 21.6 | RC circuits - charging and discharging | https://www.youtube.com/watch?v=Fgpc7z7ZA94&list=PL6oBCjihmcBmAsb_4rkKz5FKEUSJD4M4I&index=22&t=0s |
| 22.2 | Ferromagnets and Electromagnets | https://www.youtube.com/watch?v=ek7GQKVRsDE&list=PL6oBCjihmcBmAsb_4rkKz5FKEUSJD4M4I&index=244&t=0s |
| 22.2 | Ferromagnets and Electromagnets | https://www.youtube.com/watch?v=WWFeRH1Zu0M&list=PL6oBCjihmcBmAsb_4rkKz5FKEUSJD4M4I&index=78&t=0s |
| 22.2 | Electromagnetism and Ferromagnets | https://www.youtube.com/watch?v=UEK61Y7w7xE&list=PL6oBCjihmcBmAsb_4rkKz5FKEUSJD4M4I&index=111&t=0s |
| 22.2 | Ferromagnets and Electromagnets | https://www.youtube.com/watch?v=9UgIpviIOdY&list=PL6oBCjihmcBmAsb_4rkKz5FKEUSJD4M4I&index=240 |
| 22.2 | Ferromagnets & Electromagnets | https://www.youtube.com/watch?v=OUsmfqLeIfY&list=PL6oBCjihmcBmAsb_4rkKz5FKEUSJD4M4I&index=182&t=0s |
| 22.3 | Magnetic fields and field lines | https://www.youtube.com/watch?v=sBqN7_X8OLY&list=PL6oBCjihmcBmAsb_4rkKz5FKEUSJD4M4I&index=185&t=0s |
| 22.3 | Magnetic Fields and Fieldlines | https://www.youtube.com/watch?v=cxTSPCUGV8s&list=PL6oBCjihmcBmAsb_4rkKz5FKE |
| 22.3 | Magnetic Fields and Field Lines | https://www.youtube.com/watch?v=d2klj4ZyqTU&list=PL6oBCjihmcBmAsb_4rkKz5FKEUSJD4M4I&index=169&t=0s |
| 22.3 | MAGNETIC FIELDS AND FIELD LINES | https://www.youtube.com/watch?v=RmjEQBgLThw&list=PL6oBCjihmcBmAsb_4rkKz5FKEUSJD4M4I&index=240&t=0s |
| 22.4 | Lorentz Force and Right-Hand Rule One | https://www.youtube.com/watch?v=oxL9A6B6r0A&list=PL6oBCjihmcBmAsb_4rkKz5FKEUSJD4M4I&index=39&t=0s |
| 22.4 | Lorentz Force | https://www.youtube.com/watch?v=QwRUKhJ6KAg&list=PL6oBCjihmcBmAsb_4rkKz5FKEUSJD4M4I&index=178&t=0s |
| 22.4 | Lorentz Force Right Hand Rule | https://www.youtube.com/watch?v=CE8Xgv8ZZvI&list=PL6oBCjihmcBmAsb_4rkKz5FKEUSJD4M4I&index=146&t=0s |
| 22.4 | Lorentz Force and Right Hand Rule | https://www.youtube.com/watch?v=i5p3b2KjtCg&list=PL6oBCjihmcBmAsb_4rkKz5FKEUSJD4M4I&index=229&t=0s |
| 22 | AURORA BOREALIS | https://www.youtube.com/watch?v=Z_atGFldoBk&list=PL6oBCjihmcBmAsb_4rkKz5FKEUSJD4M4I&index=124&t=0s |
| 22 | Aurora Borealis | https://www.youtube.com/watch?v=sUfLFnnDhdU |
| 22 | How an Aurora Borealis works | https://www.youtube.com/watch?v=8f_F7cO5f24&list=PL6oBCjihmcBmAsb_4rkKz5FKEUSJD4M4I&index=191 |
| 22.6 | Hall Effect | https://www.youtube.com/watch?v=vLyprv9wdxc&list=PL6oBCjihmcBmAsb_4rkKz5FKEUSJD4M4I&index=135&t=0s |
| 22.6 | Hall Effect | https://www.youtube.com/watch?v=6odzj7KmkxI&list=PL6oBCjihmcBmAsb_4rkKz5FKEUSJD4M4I&index=108&t=0s |
| 22.6 | THE HALL EFFECT | https://www.youtube.com/watch?v=AgghNXj_RcU&list=PL6oBCjihmcBmAsb_4rkKz5FKEUSJD4M4I&index=60&t=0s |
| 22.6 | The Hall Effect | https://www.youtube.com/watch?v=QUc8Q7wBWmg&list=PL6oBCjihmcBmAsb_4rkKz5FKEUSJD4M4I&index=50 |
| 22.8 | Torque On A Current Loop | https://www.youtube.com/watch?v=x5gj8R5Sj9w&list=PL6oBCjihmcBmAsb_4rkKz5FKEUSJD4M4I&index=132&t=0s |
| 22.8 | Magnetic Torque On Current Loops | https://www.youtube.com/watch?v=TW317amnTWA&list=PL6oBCjihmcBmAsb_4rkKz5FKEUSJD4M4I&index=30&t=0s |
| 22.8 | Magnetic Torque | https://www.youtube.com/watch?v=n9ngRxnlglU&list=PL6oBCjihmcBmAsb_4rkKz5FKEUSJD4M4I&index=93&t=0s |
| 23.2 | FARADAYS LAW | https://www.youtube.com/watch?v=v6ToG8Xuxsw&list=PL6oBCjihmcBmAsb_4rkKz5FKEUSJD4M4I&index=345&t=0s |
| 23.2 | Magnetic Induction - Faraday's Law | https://www.youtube.com/watch?v=nq0W3SzSfic&list=PL6oBCjihmcBmAsb_4rkKz5FKEUSJD4M4I&index=346&t=0s |
| 23.2 | Faraday's Law and Induced Current | https://www.youtube.com/watch?v=r-W1Anho7Xk&list=PL6oBCjihmcBmAsb_4rkKz5FKEUSJD4M4I&index=95&t=0s |
| 23.2 | Faraday's Law | https://www.youtube.com/watch?v=kEkItL5DoxI&list=PL6oBCjihmcBmAsb_4rkKz5FKEUSJD4M4I&index=299&t=0s |
| 23.7 | Ho Transformers work | https://www.youtube.com/watch?v=YtRyuW9lkOo&list=PL6oBCjihmcBmAsb_4rkKz5FKE |
| 23.7 | Transformers | https://www.youtube.com/watch?v=VWrbTaKelds&list=PL6oBCjihmcBmAsb_4rkKz5FKEUSJD4 |
| 23.7 | TRANSFORMERS | https://www.youtube.com/watch?v=ulpbivnaeUU&list=PL6oBCjihmcBmAsb_4rkKz5FKE |
| 23.7 | Transformers | https://www.youtube.com/watch?v=HWdU8sMq5Zo&list=PL6oBCjihmcBmAsb_4rkKz5FKEUSJD4M4I&index=138&t=0s |
| 23.7 | Transformers | https://www.youtube.com/watch?v=3Egf7btZzYY&list=PL6oBCjihmcBmAsb_4rkKz5FKEUSJD4M4I&index=88&t=0s |
| 23.9 | Inductance | https://www.youtube.com/watch?v=c0qvUN9SaoE&list=PL6oBCjihmcBmAsb_4rkKz5FKEUSJD4M4I&index=139&t=0s |
| 23.9 | Rhen_Davis_Inductance..mov | https://www.youtube.com/watch?v=XUg4PzXNT10&list=PL6oBCjihmcBmAsb_4rkKz5FK |
| 23.9 | Inductance | https://www.youtube.com/watch?v=W8kaZCS3snY&list=PLYEsARTQM_GNqXYaSZx_Ks |
| 23:10 | RL Circuits | https://www.youtube.com/watch?v=zCv43CSaxSE&list=PL6oBCjihmcBmAsb_4rkKz5FKEUSJD4M4I&index=82&t=0s |
| 23:10 | RL Circuits | https://www.youtube.com/watch?v=uVyrPrdnPj4&list=PL6oBCjihmcBmAsb_4rkKz5FKEUSJD4M4I&index=24&t=0s |
| 23:10 | LR Circuits | https://www.youtube.com/watch?v=cKwKjqnnPP8&list=PL6oBCjihmcBmAsb_4rkKz5 |
| 23:12 | RLC circuits | https://www.youtube.com/watch?v=Xq-edHFc49w&list=PL6oBCjihmcBmAsb_4rkKz5FKEUSJD4M4I&index=104 |
| 23:12 | Resonance In RLC Circuits | https://www.youtube.com/watch?v=O-A89C0ywbU&list=PL6oBCjihmcBmAsb_4rkKz5FKEUSJD4M4I&index=195&t=0s |
| 23:12 | Resonance in RLC Circuits | https://www.youtube.com/watch?v=cp8YUIVJ3Cw&list=PL6oBCjihmcBmAsb_4rkKz5FKEUSJD4 |
| 23:12 | RLC CIRCUITS | https://www.youtube.com/watch?v=E91TD-xeN6Q&list=PL6oBCjihmcBmAsb_4rkKz5FKEUSJD4M4I&index=285&t=0s |
| 23:12 | Electromagnetic Waves | https://www.youtube.com/watch?v=NPImuCTneTE&list=PL6oBCjihmcBmAsb_4rkKz5FK |
| 24 | Electromagnetic Waves | https://www.youtube.com/watch?v=qG98x4ziE6M&list=PL6oBCjihmcBmAsb_4rkKz5FKEUSJD4M4I&index=168&t=0s |
| 24 | Electromagnetic Waves | https://www.youtube.com/watch?v=JCqqehjxBOo&list=PL6oBCjihmcBmAsb_4rkKz5FKEUSJD4M4I&index=32 |
| 24 | Electromagnetic Radiation | https://www.youtube.com/watch?v=hx28AkGlEGY&list=PL6oBCjihmcBmAsb_4rkKz5FKEUSJD4M4I&index=215&t=0s |
| 24 | Electromagnetic Waves | https://www.youtube.com/watch?v=idP1BFE6Hy0&list=PL6oBCjihmcBmAsb_4rkKz5FKE |
| 24.3 | Electromagnetic Spectrum | https://www.youtube.com/watch?v=VxG1pqOnhm8&list=PL6oBCjihmcBmAsb_4rkKz5FKEUSJD4M4I&index=189&t=0s |
| 24.3 | The Electromagnetic Spectrum | https://www.youtube.com/watch?v=iYwQMfiOqhM&list=PL6oBCjihmcBmAsb_4rkKz5FK |
| 24.3 | ELECTROMAGNETIC SPECTRUM | https://www.youtube.com/watch?v=mivtsaWy3WY&list=PL6oBCjihmcBmAsb_4rkKz5FKEUSJD4M4I&index=232&t=0s |
| 25.3 | Snells Law | https://www.youtube.com/watch?v=QWeuwFU1bGw&list=PL6oBCjihmcBmAsb_4rkKz5FKEUSJD4M4I&index=29&t=0s |
| 25.3 | Snells Law | https://www.youtube.com/watch?v=-PsSoUF8Bcc&list=PL6oBCjihmcBmAsb_4rkKz5FKEUSJD4M4I&index=104&t=0s |
| 25.3 | Snell's Law | https://www.youtube.com/watch?v=fP2L8kel3y8&list=PL6oBCjihmcBmAsb_4rkKz5FKEUSJD4M4I&index=246&t=0s |
| 25.3 | Snells Law | https://www.youtube.com/watch?v=VRexGsMcLcs&list=PL6oBCjihmcBmAsb_4rkKz |
| 25.2-25.3 | Reflection & Refraction | https://www.youtube.com/watch?v=5_k1lRr7UZk&list=PL6oBCjihmcBmAsb_4rkKz5FKEUSJD4M4I&index=144&t=0s |
| 25.2-25.3 | Reflection & Refraction | https://www.youtube.com/watch?v=fog6BgNv8cM&list=PL6oBCjihmcBmAsb_4rkKz5FKEUSJD4M4I&index=343&t=0s |
| 25.2-25.3 | Reflection vs Refraction | https://www.youtube.com/watch?v=ck5T9e2XKGA&list=PL6oBCjihmcBmAsb_4rkKz5FKEUSJD4M4I&index=128&t=0s |
| 25.5 | Dispersion And Rainbows | https://www.youtube.com/watch?v=3506uxuLzTM&list=PL6oBCjihmcBmAsb_4rkKz5FKEUSJD4M4I&index=293&t=0s |
| 25.5 | Dispersion and Rainbows | https://www.youtube.com/watch?v=Fliewy8uajA&list=PL6oBCjihmcBmAsb_4rkKz5FKEUSJD4M4I&index=332&t=0s |
| 25.5 | Dispersion & Rainbows | https://www.youtube.com/watch?v=qDKpuP0rRBo&list=PL6oBCjihmcBmAsb_4rkKz5FKEUSJD4M4I&index=173 |
| 25.5 | Dispersion and Rainbows | https://www.youtube.com/watch?v=zCRfssr_jy8&list=PL6oBCjihmcBmAsb_4rkKz5FKEUSJD4M4I&index=318 |
| 25.5 | DISPERSION AND RAINBOWS | https://www.youtube.com/watch?v=EuqSEFi7r1I&list=PL6oBCjihmcBmAsb_4rkKz5FKEUSJD4M4I&index=19 |
| 25.6 | Converging And Diverging Lenses | https://www.youtube.com/watch?v=5EmFNB-szrk&list=PL6oBCjihmcBmAsb_4rkKz5FKEUSJD4M4I&index=36&t=0s |
| 25.6 | CONVERGING AND DIVERGING LENSES. | https://www.youtube.com/watch?v=X50HAiL1kro |
| 25.6 | Convergin And Diverging Lenses | https://www.youtube.com/watch?v=5UbqA78Cemo&list=PL6oBCjihmcBmAsb_4rkKz5FKEUSJD4M4I&index=36&t=0s |
| 25.6 | Converging And Divering Lenses | https://www.youtube.com/watch?v=sG0O0sbgRdg&list=PL6oBCjihmcBmAsb_4rkKz5FKEUSJD4M4I&index=305&t=0s |
| 25.6 | Image Formation By Lenses | https://www.youtube.com/watch?v=AyDdkxfArT4&list=PL6oBCjihmcBmAsb_4rkKz5FKEUSJD4M4I&index=161&t=0s |
| 25.6 | Image Formation by Lenses | https://www.youtube.com/watch?v=Me9yCNmBGkY&list=PL6oBCjihmcBmAsb_4rkKz5FKEUSJD4M4I&index=206&t=0s |
| 25.6 | Lens Formation | https://www.youtube.com/watch?v=CKFFwun4slY&list=PL6oBCjihmcBmAsb_4rkKz5FKEUSJD4M4I&index=125&t=0s |
| 25.6 | Lens-Image-Formation | https://www.youtube.com/watch?v=na0R7-WPjhw&list=PL6oBCjihmcBmAsb_4rkKz5FKEUSJD4M4I&index=333&t=0s |
| 26 | Image Formation By The Eye | https://www.youtube.com/watch?v=JC6WdVItv00&list=PL6oBCjihmcBmAsb_4rkKz5FKEUSJD4M4I&index=297&t=0s |
| 26 | Image formation & Image deficiencies | https://www.youtube.com/watch?v=HRKYiK6kdDI&list=PL6oBCjihmcBmAsb_4rkKz5FKEUSJD4M4I&index=306 |
| 26 | Image Formation and Vision Deficiencies | https://www.youtube.com/watch?v=_uozYsWRBXw&list=PL6oBCjihmcBmAsb_4rkKz5FKEUSJD4M4I&index=181&t=0s |
| 26 | Image formation by the eye - vision deficiencies | https://www.youtube.com/watch?v=bKCEBwrvCU4&list=PL6oBCjihmcBmAsb_4rkKz5FK |
| 26 | Image Formation by the Eye | https://www.youtube.com/watch?v=nd8JLhP9nbU&list=PL6oBCjihmcBmAsb_4rkKz5FKEUSJD4M4I&index=114 |
| 27.2 | Huygen's_Principle | https://www.youtube.com/watch?v=iUB3r0vGLiU&list=PL6oBCjihmcBmAsb_4rkKz5FKEUSJD4M4I&index=106&t=0s |
| 27.2 | Huygens Principle | https://www.youtube.com/watch?v=dULsn8Jawmk&list=PL6oBCjihmcBmAsb_4rkKz5FKEUS |
| 27.2 | Huygens Principle | https://www.youtube.com/watch?v=kczaM2ogRGQ&list=PL6oBCjihmcBmAsb_4rkKz5FKEUSJD4M4I&index=179&t=0s |
| 27.2 | HUYGENS PRINCIPLE | https://www.youtube.com/watch?v=vsrgWuqTfgQ&list=PL6oBCjihmcBmAsb_4rkKz5FKEUSJD4M4I&index=54&t=0s |
| 27.2 | HUYGENS PRINCIPLE | https://www.youtube.com/watch?v=Z94HT-csHvE&list=PL6oBCjihmcBmAsb_4rkKz5FKEUSJD4M4I&index=98&t=0s |
| 27.3 | Constructive & Destructive Interference | https://www.youtube.com/watch?v=_WMiPI49C8A&list=PL6oBCjihmcBmAsb_4rkKz5FKEUSJD4M4I&index=272&t=0s |
| 27.3 | Constructive and Destructive Interference | https://www.youtube.com/watch?v=0bbN55XoaAk&list=PL6oBCjihmcBmAsb_4rkKz5FKEUSJD4M4I&index=294&t=0s |
| 27.3 | Constructive and destructive interference | https://www.youtube.com/watch?v=N5GF4QLj9sE&list=PL6oBCjihmcBmAsb_4rkKz5FKE |
| 27.3 | Constructive and destructive interference | https://www.youtube.com/watch?v=AMXn0O8sTYw&list=PL6oBCjihmcBmAsb_4rkKz5FKEUSJD4M4I&index=82&t=0s |
| 27.3 | Constructive And Destructive Interference | https://www.youtube.com/watch?v=gXfleDcfC18&list=PL6oBCjihmcBmAsb_4rkKz5FKEU |
| 27.6 | Rayleigh Criterion | https://www.youtube.com/watch?v=60TAhLw-UjM&list=PL6oBCjihmcBmAsb_4rkKz5FKEUSJD4M4I&index=86&t=0s |
| 27.6 | Rayleighs Criterion | https://www.youtube.com/watch?v=YSlqrcvmBAE&list=PL6oBCjihmcBmAsb_4rkKz5FKE |
| 27.6 | Rayleigh Criterion | https://www.youtube.com/watch?v=gADPBqZQii0&list=PL6oBCjihmcBmAsb_4rkKz5FKEUSJD4M4I&index=184&t=0s |
| 27.8 | POLARIZATION | https://www.youtube.com/watch?v=O3zq8R03aQE&list=PL6oBCjihmcBmAsb_4rkKz5FKEUSJD4M4I&index=186 |
| 27.8 | Polarization | https://www.youtube.com/watch?v=yCfF1KLMQCA&list=PL6oBCjihmcBmAsb_4rkKz |
| 28.1 | Einstein's Postulates | https://www.youtube.com/watch?v=GUeSfI1u30g&list=PL6oBCjihmcBmAsb_4rkKz5FKEUSJD4M4I&index=233 |
| 28.1 | Time Dilation and Length Contraction | https://www.youtube.com/watch?v=UusFjC2UifA&list=PL6oBCjihmcBmAsb_4rkKz5FKEUSJD4M4I&index=23&t=0s |
| 28.1 | Two postulates of Special Relativity | https://www.youtube.com/watch?v=Di83eWMHwB8&list=PL6oBCjihmcBmAsb_4rkKz5FKEUSJD4M4I&index=78&t=0s |
| 28.1 | Two Postulates of Special Relativity | https://www.youtube.com/watch?v=3dTV9M-0wpk&list=PL6oBCjihmcBmAsb_4rkKz5FKEUSJD4M4I&index=71&t=0s |
| 28.2-28.3 | Length contraction and time dilation | https://www.youtube.com/watch?v=e_JAMZ-tpiE&list=PL6oBCjihmcBmAsb_4rkKz5FKEUSJD4M4I&index=29&t=0s |
| 28.2-28.3 | Time Dilation and Length Contraction | https://www.youtube.com/watch?v=7cIXl2t0Y4U&list=PL6oBCjihmcBmAsb_4rkKz5FKEUSJD4M4I&index=70&t=0s |
| 28.2-28.3 | Length Contraction & Time Dilation | https://www.youtube.com/watch?v=xMYEArwZPfs&list=PL6oBCjihmcBmAsb_4r |
| 28.2-28.3 | Time and Length Dilation | https://www.youtube.com/watch?v=CZpqiePqIXU&list=PL6oBCjihmcBmAsb_4rkKz5FKEUSJD4M4I&index=94&t=0s |
| 28.2-28.3 | Time and Length Dilation And Contraction | https://www.youtube.com/watch?v=uVnbhbFdbwI&list=PL6oBCjihmcBmAsb_4r |
| 28.4 | Relativistic addition of velocities | https://www.youtube.com/watch?v=IQCYsrkhBO8&list=PL6oBCjihmcBmAsb_4rkKz5FKE |
| 28.4 | Dopplershift | https://www.youtube.com/watch?v=HoD_XYxewII&feature=youtu.be |
| 28.4 | Doppler Shift | https://www.youtube.com/watch?v=8V09Uc6HgTA&list=PL6oBCjihmcBmAsb_4rkKz5FKEUSJD4M4I&index=190&t=0s |
| 28.4 | Doppler Shift | https://www.youtube.com/watch?v=_jhJSXI8Pxc&list=PL6oBCjihmcBmAsb_4rkKz5FKEUSJD4M4I&index=209&t=0s |
| 28.4 | Doppler Shift | https://www.youtube.com/watch?v=A_IRz8DBK_8&list=PL6oBCjihmcBmAsb_4rkKz5FKEUSJD4M4I&index=131 |
| 28.5-28.6 | Relativistic Momentum and Energy | https://www.youtube.com/watch?v=s8UF346f1xE&list=PL6oBCjihmcBmAsb_4rkKz5FKE |
| 28.5-28.6 | Relativistic Momentum | https://www.youtube.com/watch?v=O25gv4101xA&list=PL6oBCjihmcBmAsb_4rkKz5FKEUSJD4M4I&index=221&t=0s |
| 28.5-28.6 | Relative Momentum & Energy | https://www.youtube.com/watch?v=TxwbN1KRhj0&list=PL6oBCjihmcBmAsb_4rkKz5FKEUSJD4M4I&index=170&t=0s |
| 29.1 | Blackbody Radiation Evidence For Quantum Theory | https://www.youtube.com/watch?v=ULP72z-obyc&list=PL6oBCjihmcBmAsb_4rkKz5FKEUSJD4M4I&index=61&t=0s |
| 29.1 | Blackbody Radiation as Evidence for Quantum Mechanics | https://www.youtube.com/watch?v=t_actghwRQE&list=PL6oBCjihmcBmAsb_4rkKz5FKEUSJ |
| 29.1 | Blackbody radiation | https://www.youtube.com/watch?v=S7Tp3E0L-3U&list=PL6oBCjihmcBmAsb_4rkKz5FKEUSJD4M4I&index=58&t=0s |
| 29.1 | Atomic spectra | https://www.youtube.com/watch?v=YNiEHU1obQA&list=PL6oBCjihmcBmAsb_4rkKz5FKEUSJD4M4I&index=303&t=0 |
| 29.1 | Atomic Spectra | https://www.youtube.com/watch?v=pzo10qYg1YU&list=PL6oBCjihmcBmAsb_4rkKz5FKEUSJD4 |
| 29.1 | Atomic Spectra | https://www.youtube.com/watch?v=7XRwumVz8Bk&list=PL6oBCjihmcBmAsb_4rkKz5FKEUSJD4M4I&index=2&t=0s |
| 29.1 | Atomic Spectra | https://www.youtube.com/watch?v=pzo10qYg1YU&list=PL6oBCjihmcBmAsb_4rkKz5FKEUSJD4M4I&index=159&t=0s |
| 29.1 | Atomic spectra | https://www.youtube.com/watch?v=7XRwumVz8Bk&list=PL6oBCjihmcBmAsb_4rkKz5FKEUSJD4M4I&index=2&t=0s |
| 29.2 | Photoelectric Effect | https://www.youtube.com/watch?v=Jy1YCZN0b80&list=PL6oBCjihmcBmAsb_4rkKz5FKEUSJD4M4I&index=103&t=0s |
| 29.2 | Photoelectric Effect | https://www.youtube.com/watch?v=sLz809gwP-w&list=PL6oBCjihmcBmAsb_4rkKz5FKEUSJD4M4I&index=66&t=0s |
| 29.2 | Photoelectric Effect | https://www.youtube.com/watch?v=RVuqHCbCw2c&list=PL6oBCjihmcBmAsb_4rkKz5FKEUSJD4M4I&index=127&t=0s |
| 29.2 | The Photoelectric Effect | https://www.youtube.com/watch?v=ehBu-c8O_g0&list=PL6oBCjihmcBmAsb_4rkKz5FKEUSJD4M4I&index=118&t=0s |
| 29.4 | Applications of Photon Momentum | https://www.youtube.com/watch?v=KIx7nz8XAIg&list=PL6oBCjihmcBmAsb_4rkKz5FKEUSJD4M4I&index=166&t=12s |
| 29.4 | Photon Momentum Applications | https://www.youtube.com/watch?v=l-48k6fpnw8&list=PL6oBCjihmcBmAsb_4rkKz5FKEUSJD4M4I&index=182&t=0s |
| 29.4 | Photon Momentum (Physics 106 Dr. Della Corte) | https://www.youtube.com/watch?v=r3PswZRvY5U&feature=youtu.be&hd=1 |
| 29.4 | Photon Momentum | https://www.youtube.com/watch?v=QuJ1FOBVpoE&list=PL6oBCjihmcBmAsb_4r%20kKz5FKEUSJD4M4I&index=100&t=0s |
| 29.4 | Photon Momentum Applications | https://www.youtube.com/watch?v=EiK10kUbTl8&list=PL6oBCjihmcBmAsb_4rkKz5FKEUSJD4M4I&index=48 |
| 29.5 | The Wave-Particle Duality of Photons | https://www.youtube.com/watch?v=TH8TA7-JGVc |
| 29.5 | Duality Of Photons | https://www.youtube.com/watch?v=0D_3Pmr2hW0&list=PL6oBCjihmcBmAsb_4rkKz5F |
| 29.5 | Wave Particle Duality | https://www.youtube.com/watch?v=M55P6X2rHtU&list=PL6oBCjihmcBmAsb_4rkKz5FKEUSJD4M4I&index=137&t=0s |
| 29.5 | Wave particle duality of photons | https://www.youtube.com/watch?v=DrXw6X9W6Nw&list=PL6oBCjihmcBmAsb_4rkKz5FKEUSJD4M4I&index=162&t=0s |
| 29.7 | Heisenberg uncertainty principle | https://www.youtube.com/watch?v=KfFuJadvSGs&list=PL6oBCjihmcBmAsb_4rkKz5FKE |
| 29.7 | Heisenberg Uncertainty | https://www.youtube.com/watch?v=ekkW8TkDUys&list=PL6oBCjihmcBmAsb_4r%20kKz5FKEUSJD4M4I&index=197&t=0s |
| 29.7 | Heisenbergs Uncertainty Principle | https://www.youtube.com/watch?v=ynfm7BzfM0o&list=PL6oBCjihmcBmAsb_4rkKz5FK |
| 29.7 | Heisenberg Uncertainty Principle | https://www.youtube.com/watch?v=cI0taASnshQ&list=PL6oBCjihmcBmAsb_4rkKz5FKEUSJD4M4I&index=203&t=0s |
| 30.2-30.3 | Rutherfold and Bohrmodel of the atom | https://www.youtube.com/watch?v=uv5GKPYo9Eo&list=PL6oBCjihmcBmAsb_4rkKz5FKEUSJD4M4I&index=101&t=0s |
| 30.2-30.3 | Bohr Atomic Models | https://www.youtube.com/watch?v=-aVpt_l2uXs&list=PL6oBCjihmcBmAsb_4rkKz5FKEUSJD4M4I&index=218 |
| 30.2-30.3 | Rutherford And Bohr Model Of The Atom | https://www.youtube.com/watch?time_continue=1&v=9PUPkHjkyC8&feature=emb_logo |
| 30.2-30.3 | RUTHERFORD & BOHR MODEL | https://www.youtube.com/watch?v=A10FtGjUm6I&list=PL6oBCjihmcBmAsb_4rkKz5FKEUSJD4M4I&index=343&t=0s |
| 30.2-30.3 | Bohr-Rutherford Model | https://www.youtube.com/watch?v=JmTFeaeTRs4&list=PL6oBCjihmcBmAsb_4rkKz5FKEUSJD4M4I&index=52&t=0s |
| 30.5 | Lasers | https://www.youtube.com/watch?v=ByYFCT93IaQ&list=PL6oBCjihmcBmAsb_4rkKz5FKEUSJD4M4I&index=321 |
| 30.5 | How Laser work | https://www.youtube.com/watch?v=VGCVZ_hREXA&list=PL6oBCjihmcBmAsb_4rkKz5FKEUSJD4M4I&index=26&t=0s |
| 30.5 | How Lasers Work | https://www.youtube.com/watch?v=cDf_rNjYf0g&list=PL6oBCjihmcBmAsb_4rkKz5FKEUSJD%204M4I&index=343&t=0s |
| 30.5 | Lasers | https://www.youtube.com/watch?v=nzs8Xl0MiqY&list=PL6oBCjihmcBmAsb_4rkKz5FKEUSJD4M4I&index=248&t=0s |
| 30.7 | Zeeman Effect | https://www.youtube.com/watch?v=t2PfHgGdpOE&list=PL6oBCjihmcBmAsb_4rkKz5FKEUSJD4M4I&index=134&t=0s |
| 30.7 | The Zeeman Effect | https://www.youtube.com/watch?v=ZT1rnGFTesU&list=PL6oBCjihmcBmAsb_4rkKz5FKEUSJD4M4I&index=40 |
| 30.7 | Zeeman Effect | https://www.youtube.com/watch?v=oXxTMTCRgQA&list=PL6oBCjihmcBmAsb_4%20rkKz5FKEUSJD4M4I&index=99&t=0s |
| 30.7 | Zeeman Effect | https://www.youtube.com/watch?v=MUNuZYE4COw&list=PL6oBCjihmcBmAsb_4rkKz |
| 30.8 | Quantum Numbers | https://www.youtube.com/watch?v=Vq6D9l7sxpw&list=PL6oBCjihmcBmAsb_4rkKz5FKEUSJD4M4I&index=128&t=0s |
| 30.8 | Quantum Numbers Electron Spin | https://www.youtube.com/watch?v=MqB6DNdUFmc&list=PL6oBCjihmcBmAsb_4rkKz5FKEUSJD4M4I&index=187&t=0s |
| 30.8 | Electron spin quantum numbers | https://www.youtube.com/watch?v=m3MSk1e9coE |
| 30.8 | Quantum Numbers and Electron Spin | https://www.youtube.com/watch?v=Bfp1qgMplS0 |
| 30.9 | Pauli Exclusion Principle | https://www.youtube.com/watch?v=iUaUrN4y88w&list=PL6oBCjihmcBmAsb_4rkKz5FKEUS%20JD4M4I&index=341&t=240s |
| 30.9 | Pauli Exclusion Principle | https://www.youtube.com/watch?v=sOe32I8DD8k&list=PL6oBCjihmcBmAsb_4rkKz5FKEUSJD4%20M4I&index=85&t=0s |
| 30.9 | Pauli Exclusion Principle | https://www.youtube.com/watch?v=2EV1MEDH06Q&list=PL6oBCjihmcBmAsb_%204rkKz5FKEUSJD4M4I&index=37&t=0s |
| 30.9 | Pauli Exclusion Principle | https://www.youtube.com/watch?v=1RA8LbRJOXQ&list=PL6oBCjihmcBmAsb_4rkKz5FKEUSJD4M4I&index=301&t=0s |
| 30.9 | Pauli Exclusion Principle | https://www.youtube.com/watch?v=E4QzjuNsl90&list=PL6oBCjihmcBmAsb_4rkKz5FKEUSJD4M4I&index=46&t=0s |
| 33.2 | Four Fundamental Forces | https://www.youtube.com/watch?v=JQxV7zPfctg&list=PL6oBCjihmcBmAsb_4rkKz5FKEUSJD4M%204I&index=245&t=0s |
| 33.2 | Fundamental Forces | https://www.youtube.com/watch?v=TW230tN5ykU&list=PL6oBCjihmcBmAsb_4rkKz5FKEUSJD4M4I&index=312&t=0s |
| 33.2 | Four Fundamental Forces | https://www.youtube.com/watch?v=OWQHKsJ4eGs&list=PL6oBCjihmcBmAsb_4rkKz5FKEUSJD4M4I&index=309&t=0s |
| 33.2 | Four Fundamental Forces | https://www.youtube.com/watch?v=681NFLWkL04&list=PL6oBCjihmcBmAsb_4rkKz5FKEUSJD4M4I&index=320 |
| 31.3 | Structure of Nucleus | https://www.youtube.com/watch?v=GNDighwWUz4&list=PL6oBCjihmcBmAsb_4rkKz5FKEUSJD4M4I&index=102&t=0s |
| 31.3 | Structure of the Nucleus | https://www.youtube.com/watch?v=3c3ZUvxBqlo&list=PL6oBCjihmcBmAsb_4rkKz5FKEUS%20JD4M4I&index=144&t=0s |
| 31.3 | THE NUCLEUS | https://www.youtube.com/watch?v=c7qLzWiE_tc&list=PL6oBCjihmcBmAsb_4rkKz5FKEUSJD4M4I&index=213&t=0s |
| 31.3 | Structure of the Nucleus | https://www.youtube.com/watch?v=ssKKHTa-eos&list=PL6oBCjihmcBmAsb_4rkKz5FKEUSJD4M4I&index=163&t=0s |
| 31.1 | Alpha Beta Decay | https://www.youtube.com/watch?v=xKgKGcH5a5E&list=PL6oBCjihmcBmAsb_4rkKz5FKEUSJD4%20M4I&index=310&t=0s |
| 31.1 | Radioactive Decay | https://www.youtube.com/watch?v=opUtqoCfbLI&list=PL6oBCjihmcBmAsb_4rk |
| 31.1 | Alpha, Beta, and GammaDecay | https://www.youtube.com/watch?v=0jbIFzNoKsQ&list=PL6oBCjihmcBmAsb_4rkKz5FKEUS%20JD4M4I&index=89&t=0s |
| 31 | Radiation Damage and Prevention | https://www.youtube.com/watch?v=NHKpQ4g2riE&list=PL6oBCjihmcBmAsb_4rkKz5FKE |
| 31 | Damage and Prevention of Radiation. | https://www.youtube.com/watch?v=IGeL9RyHqMw&list=PL6oBCjihmcBmAsb_4rkKz5FKEUSJD4%20M4I&index=285&t=0s |
| 31 | Radiation Damage | https://www.youtube.com/watch?v=_jmqba3Y-mU&list=PL6oBCjihmcBmAsb_4rkKz5FKEUSJD4M4I&index=113&t=0s |
| 31.5 | Carbon Dating | https://www.youtube.com/watch?v=77lAiiqHiMA&list=PL6oBCjihmcBmAsb_4rkKz5FKEUSJD4M4I&index=220&t=0s |
| 31.5 | Carbon Dating | https://www.youtube.com/watch?v=jdjFlqvRiqw&list=PL6oBCjihmcBmAsb_4rkKz5FKEUSJD%204M4I&index=337&t=0s |
| 31.5 | Carbon Dating | https://www.youtube.com/watch?v=-gZYjPVOLWA&list=PL6oBCjihmcBmAsb_4rkKz5FKEUSJD4M4I&index=154&t=0s |
| 22.5 | Effect of Magnetic Field on Mars vs Earth | https://www.youtube.com/watch?v=cOFxGSmMd9E&list=PL6oBCjihmcBmAsb_4rkKz |
| 22 | Paramagnetic Levitation | https://www.youtube.com/watch?v=sPvlgipkqL4&list=PL6oBCjihmcBmAsb_4rkKz5FKEUSJD4M4I&index=53&t=0s |
| 22 | Paramagnetism and Levitation | https://www.youtube.com/watch?v=asbzv8zevNk&list=PL6oBCjihmcBmAsb_4rkKz5FKEUSJD4M4I&index=237&t=0s |
| 22 | Magnetic Levitation | https://www.youtube.com/watch?v=9Kt_64_6vFY&list=PL6oBCjihmcBmAsb_4rkKz5FKEUSJD4M4I&index=260&t=0s |
| 22 | Diamagnetic Levitation | https://www.youtube.com/watch?v=phcRadqhMqo&list=PL6oBCjihmcBmAsb_4rkKz5FKEUSJD4M4I&index=115 |
| 30.9 | Origins of the Periodic Table | https://www.youtube.com/watch?v=E5Yad3poDcs&list=PL6oBCjihmcBmAsb_4rkKz5FKEUSJD4M4I&index=239&t=0s |
| 30.9 | History Of The Periodic Table | https://www.youtube.com/watch?v=URY4bDST9FE&list=PL6oBCjihmcBmAsb_4rkKz5FK |
| 30.9 | History Of The Periodic Table | https://www.youtube.com/watch?v=voOIiv268do&list=PL6oBCjihmcBmAsb_4rkKz5FKEUSJD4M |
| 30.9 | History Of The Periodic Table | https://www.youtube.com/watch?v=2QmZEnar-Ho&list=PL6oBCjihmcBmAsb_4rkKz5FKEUSJD4M4I&index=83&t=0s |
| N/A | How does a Tesla Battery Work | https://www.youtube.com/watch?v=TuRF_B1d7dc&list=PL6oBCjihmcBmAsb_4rkKz5FKEUSJD4%20M4I&index=57&t=0s |
| N/A | How a Tesla Car Battery Works | https://www.youtube.com/watch?v=6g5pUGHvU78&list=PL6oBCjihmcBmAsb_4rkKz5FKEUSJD4M4I&index=80&t=0s |
| N/A | How does a Tesla Car battery work | https://www.youtube.com/watch?v=APdIBaMi2cc&feature=youtu.be |
| N/A | How a Tesla Car Battery works | https://www.youtube.com/watch?v=DofznS9pXY0&list=PL6oBCjihmcBmAsb_4rkKz5FKE |